\documentclass{ws-procs9x6}

\begin{document}

\title{Transverse Spin and RHIC}

\author{L.C.Bland}

\address{Brookhaven National Laboratory, \\
Upton, NY 11786, USA \\
E-mail: bland@bnl.gov}

\maketitle

\abstracts{
The Relativistic Heavy Ion Collider (RHIC) at Brookhaven
National Laboratory is the first accelerator facility that can accelerate,
store and collide spin polarized proton beams.  This development
enables a physics program aimed at understanding how the
spin of the proton results from its quark and gluon substructures.
Spin states that are either parallel (longitudinal) or perpendicular
(transverse) to the proton momentum reveal important insight into
the structure of the proton.  This talk outlines future plans for
further studies of transverse spin physics at RHIC.}

\section{Introduction}

There has been renewed experimental and
theoretical interest in transverse spin physics.  Large transverse single-spin
asymmetries (SSA) observed in elastic proton scattering and particle production
experiments (hyperon production and pion production) were often viewed
as a challenge to QCD, since the chiral properties of the theory should make
transverse single spin asymmetries small for inclusive particle
production.  Many people believed that transverse SSA would disappear
when studying polarized p+p collisions at higher collision energies ($\sqrt{s}$)
now possible at the Relativistic Heavy Ion Collider (RHIC)
at Brookhaven National Laboratory.

The modern perspective views transverse SSA
as a challenge to our understanding of hadrons on long distance
scales, possibly providing sensitivity to the transversity structure
function or to spin- and transverse-momentum dependent distribution functions
that are related to parton orbital motion.  

In this contribution, I briefly review transverse SSA and related
measurements completed at RHIC to date and describe what new 
measurements are expected in the near-term future.

\section{Recent developments in transverse spin physics}

This workshop surveyed ongoing experimental and theoretical work.  A
deeper understanding of spin- and transverse-momentum dependent
distribution functions (embodied in the Sivers effect\cite{sivers}) and
fragmentation functions (one of the keys to the Collins effect\cite{collins}) has
recently emerged.  The former were known to violate ``naive'' time
reversal symmetry.  Results from a specific model demonstrate their
possible existence\cite{BHS}.  This important theoretical development was
essentially concurrent with experimental results from semi-inclusive
deep inelastic scattering from a transversely polarized proton target
that unambiguously observed a non-zero Sivers effect\cite{HERMES}.  Also concurrent
was the observation of large spin effects in dihadron correlations
from $e^+e^-$ collisions that indicates that spin-dependent fragmentation 
effects are large.  This is one of the keys to the Collins
effect\cite{Belle}.  These experimental developments, coupled with the observation
of non-zero single spin asymmetries in inclusive pion production from
the first collisions at RHIC, have led to a reinvigoration of
transverse spin physics.

\begin{figure}[ht]
\centerline{\epsfxsize=1.9in\epsfbox{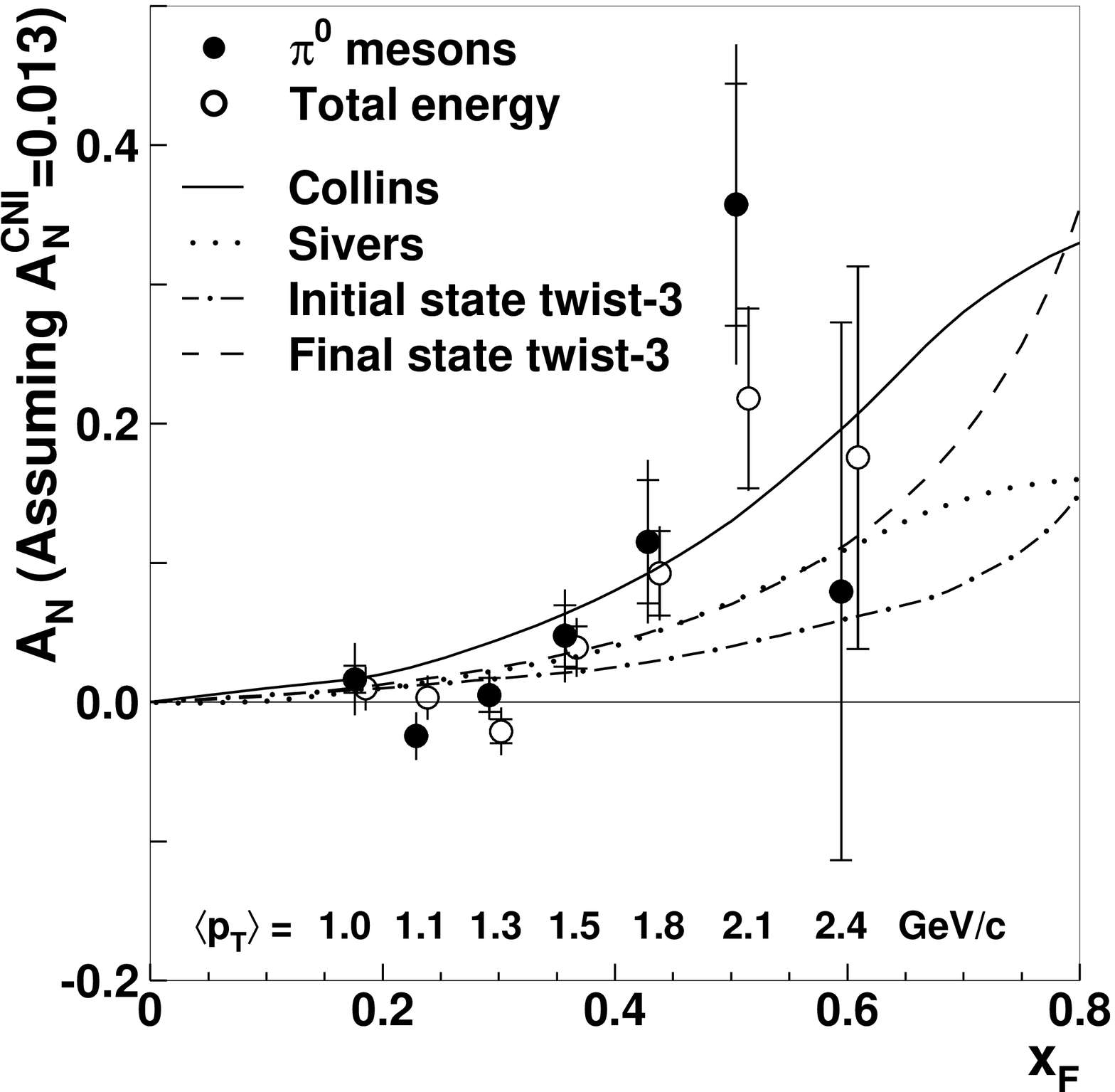}~~\epsfxsize=2.0in\epsfbox{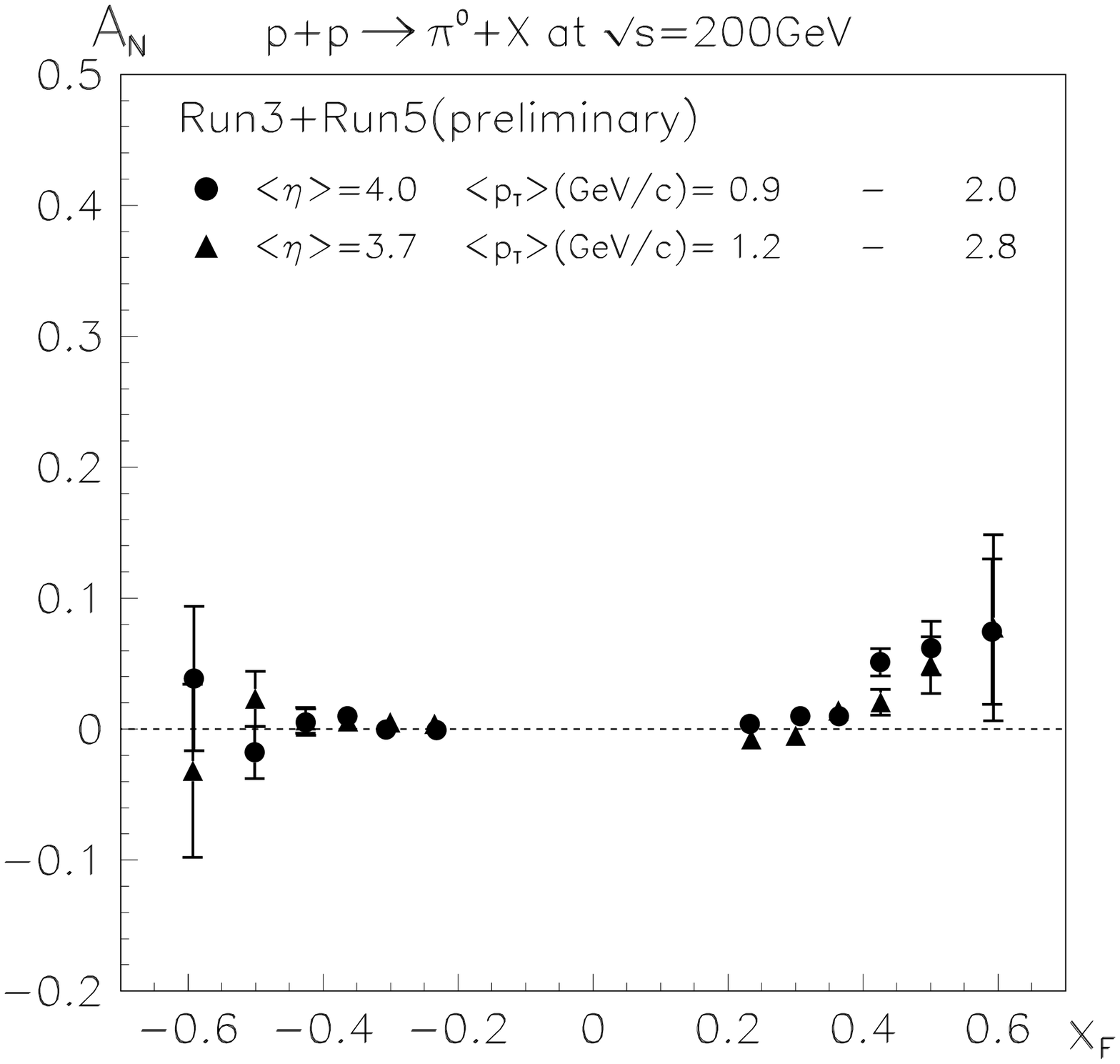}}
\caption{(left) First results for $p_{\uparrow}+p\rightarrow\pi^0+X$ analyzing powers 
  ($A_N$) at $\sqrt{s}=200$ GeV compared to theoretical predictions
  available prior to the measurements.  (right) More precise preliminary results,
  using on-line measurements of the beam polarization, for
  $\pi^0$ production $A_N$ obtained in subsequent runs.  \label{STAR_AN_results}}
\end{figure}

Large SSA were observed for $p_{\uparrow}+p\rightarrow\pi^0+X$ at
$\sqrt{s}=200$ GeV by the STAR collaboration\cite{FPD} in the first
polarized proton collisions at RHIC.  They confirmed the expectation
\cite{anselminocollins,anselminosivers,qiusterman,koike}, not shared
by all, that the sizeable SSA observed for pion production at 
$\sqrt{s}=20$ GeV\cite{E704} would persist at an order of magnitude higher collision
energy.  These expectations are shown by the theoretical predictions
in Fig. \ref{STAR_AN_results}, available prior to the measurements.
Subsequent development of full integration over intrinsic transverse
momentum has modified the relative contributions to transverse SSA
from different sources\cite{kperp}.  More recent data\cite{Morozov}
have improved the statistical precision of the effect and given the
first hint of its separated $x_F$ and $p_T$ dependence.  Preliminary
results from the BRAHMS collaboration indicate that mirror
asymmetries ($A_N(\pi^-)\approx-A_N(\pi^+)$) are observed for large
rapidity $\pi^{\pm}$ production\cite{brahms}, similar to the
lower-energy results\cite{E704}. 

\begin{figure}[ht]
\centerline{\epsfxsize=1.9in\epsfbox{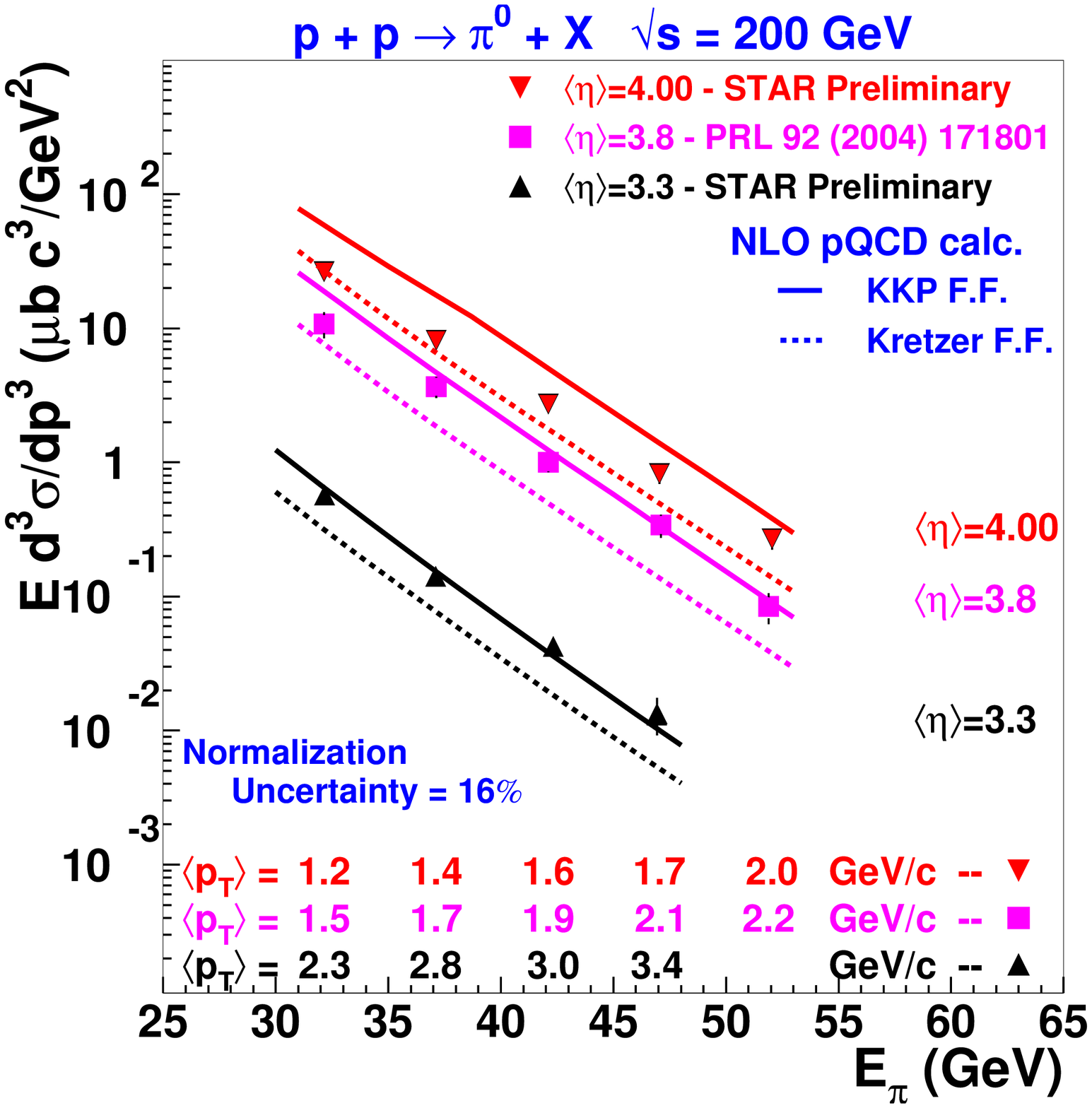}~~\epsfxsize=1.9in\epsfbox{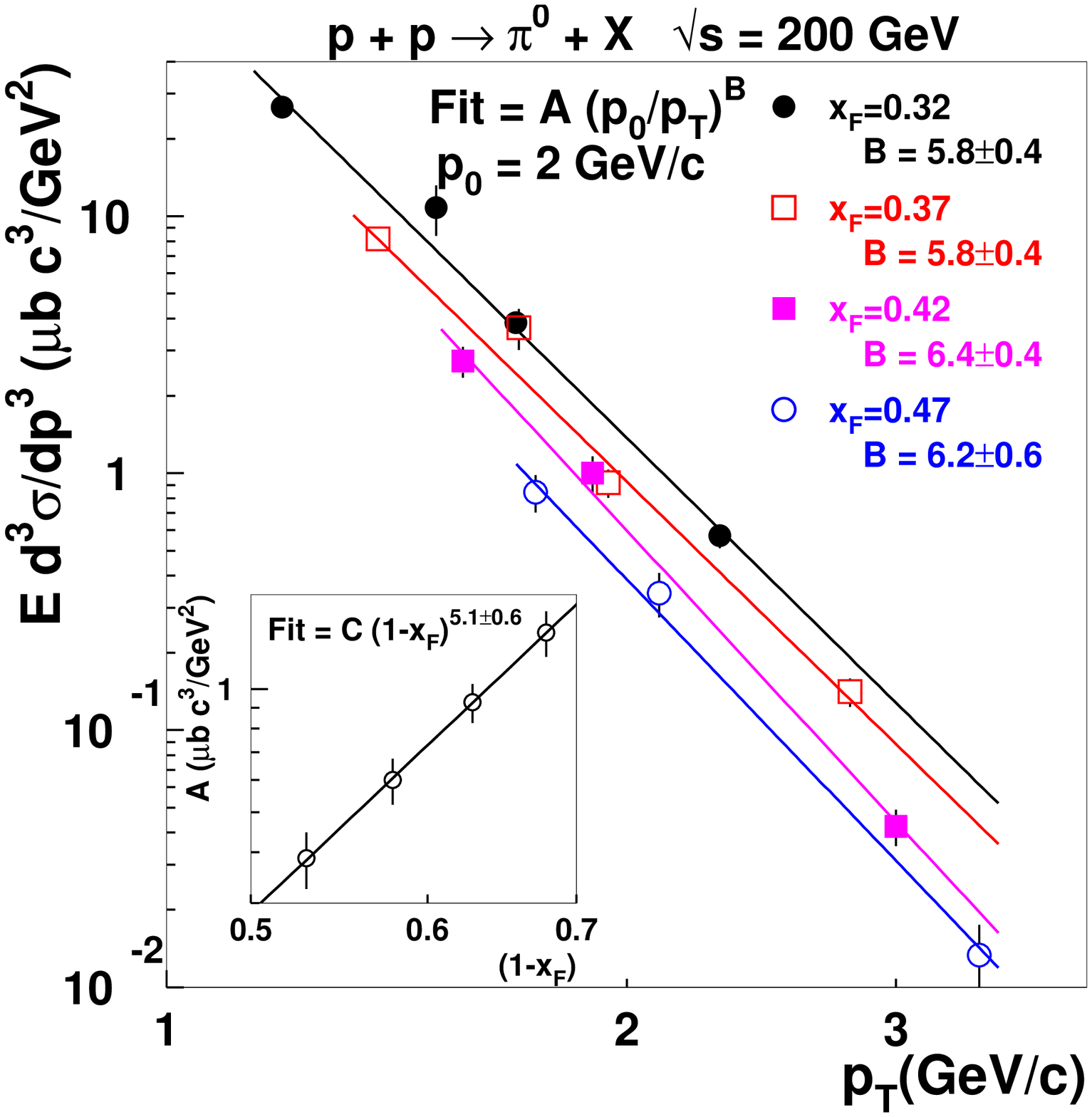}}
\caption{Results for $p+p\rightarrow\pi^0+X$ cross sections
  at $\sqrt{s}=200$ GeV compared to NLO pQCD calculations
  using conventional parton distribution and fragmentation functions.  
  (right) Parameterized $x_F$ and $p_T$ dependence.  \label{STAR_sig_results}}
\end{figure}

Perhaps most significantly, it has been established that $\pi$
production cross sections at RHIC collision energies, in the
kinematics where single spin effects are observed, are consistent with
next-to-leading order perturbative QCD (NLO pQCD) calculations at
$\sqrt{s}=200$ GeV (Fig. \ref{STAR_sig_results}).  This is in marked
contrast to the situation at lower $\sqrt{s}$ where measured cross
sections far exceed NLO pQCD predictions\cite{bs}, apparently consistent with
the belief that the transverse SSA in hadroproduction were due to beam
fragmentation.  The NLO pQCD description at $\sqrt{s}=200$ GeV
describes particle production being due to partons from both beams
undergoing a hard scattering prior to fragmenting to the observed hadrons.  That
description is further supported by experimental data that shows a significant
back-to-back peak for hadrons detected at midrapidity for events where
a large rapidity $\pi^0$ is observed\cite{Ogawa}.  

Measurements of the cross section for inclusive
$\pi^0$,\cite{PHENIX_pi0} charged hadron\cite{PHENIX_had} and jet
production\cite{STAR_jet} at midrapidity have been completed and compared with NLO
pQCD calculations\cite{QCD_calc}.  Quantitative agreement with
calculations has been found.  This agreement is an important basis for
the interpretation of spin observables ($A_N$ and the helicity
asymmetry, $A_{LL}$, that is sensitive to gluon polarization).
Transverse SSA for midrapidity $\pi^0$ and charged hadron
production\cite{PHENIX_had} have been measured and are consistent with
zero with a precision comparable to the non-zero $A_N$ found at large
rapidity at the same $p_T$ of $\sim$2 GeV/c.  The midrapidity
results may lead to important constraints on the magnitude of the
gluon Sivers function.

\section{The future}

\begin{figure}[ht]
\centerline{\epsfxsize=3.5in\epsfbox{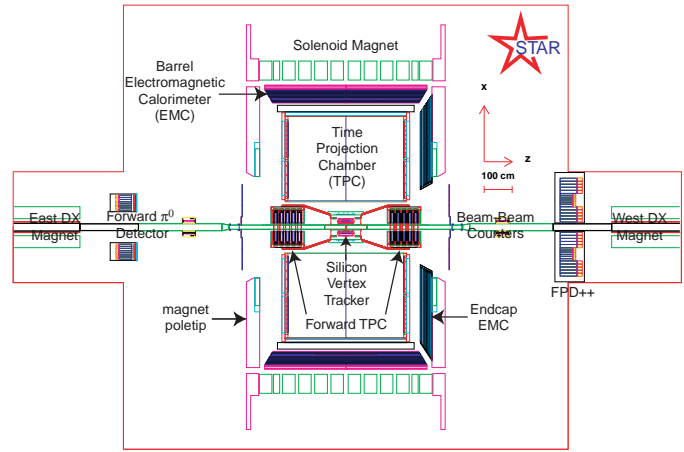}}
\caption{Layout of the STAR experiment for RHIC run 6.  The Forward $\pi^0$ Detector (FPD)
arrays shown east of the STAR magnet were also present in earlier runs west of the
STAR magnet.  For run 6, the west FPD has been upgraded to become the
FPD++.  The Forward Time Projection Chamber provides information about
charged hadrons in the angular range spanned by the forward calorimetry.\label{STAR}}
\end{figure}

\begin{figure}[ht]
\centerline{\epsfxsize=3.5in\epsfbox{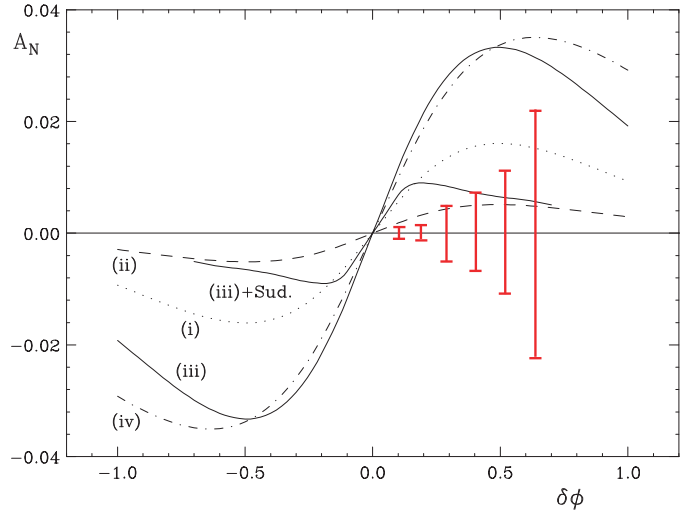}}
\caption{Projected sensitivity to the analyzing power versus the
  azimuthal angle difference between pairs of jets detected at
  midrapidity for a data sample corresponding to 5 pb$^{-1}$ with beam
  polarization of 50\% compared to theoretical expectations from
  different models of the gluon Sivers function as discussed in the
  text.\label{STAR_dijets}} 
\end{figure}

I'll restrict attention to the near-term future since there will be
significant data sets obtained with transverse polarization during the
upcoming RHIC run following the resolution of budgetary problems.  A
main objective for midrapidity studies of $p_{\uparrow}+p$ collisions
at $\sqrt{s}$=200 GeV is to establish if there are spin effects
correlated with $k_T$, a transverse momentum imbalance that is
observable if more than one particle, or more than one jet, is
observed.  Such effects could be a signal of a non-zero Sivers
function for gluons\cite{BV}.  STAR (Fig. \ref{STAR}) plans to measure $k_T$ for di-jet
events (Fig. \ref{STAR_dijets}) and will use vertical polarization.
The projected sensitivity is based on existing unpolarized data for
the azimuthal angle difference between pairs of midrapidity
jets\cite{Henry}.  PHENIX plans to measure $k_T$ by detecting pairs of
hadrons in their central arms whose symmetry requires radial
polarization to observe a spin effect.  Non-zero transverse SSA for
midrapidity dihadron production may have contributions from both the Sivers
effect and the Collins effect\cite{BBMP}. 

A portion of the upcoming RHIC run will be devoted to collisions of
transversely polarized protons at $\sqrt{s}$=62 GeV.  BRAHMS aims to
measure transverse SSA for inclusive production of identified charged
hadrons at large rapidity ($\eta\approx$3.3 and 3.9) from these
collisions.  Their particle identification apparatus will permit
measurements up to $x_F\approx$0.6 at the lower $\sqrt{s}$.  The
unpolarized cross section systematics discussed earlier\cite{bs}
would greatly benefit from new forward angle results at $\sqrt{s}$=62
GeV.  In the remainder of this section, I'll discuss plans in the
upcoming RHIC run for measurements with increased acceptance forward
calorimetry in STAR.

\begin{figure}[ht]
\centerline{\epsfxsize=2.5in\epsfbox{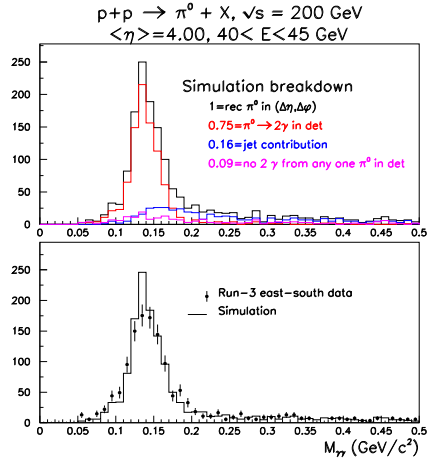}}
\caption{(bottom) Di-photon invariant mass distribution for events
with at least two reconstructed photons.  Data compares well to PYTHIA/GEANT
simulations.  (top) Contributions to simulated di-photon invariant mass
distribution for $N_{\gamma}>1$ events.  Jet-like events are evident.
\label{jet_like}}
\end{figure}

An important goal is to address the relative
contributions from the Collins and Sivers effects to the transverse
SSA observed for inclusive forward pion production.  One way to disentangle
the contributions is to address the question
``{\it is there a significant transverse single
spin asymmetry for jet-like events in p+p collisions?}''  

Jet-like events are defined as having three or more photons which are
mostly a $\pi^0$ and accompanying particles or single photon daughters from two or more
$\pi^0$.  In either case, multiple fragments of the parton scattered
through small angles are observed, making the events manifestly jet
like.  If the detector acceptance for the observed particles is
azimuthally symmetric around the thrust axis of the forward scattered
parton, then a transverse SSA for jet-like events must be due to the
Sivers effect\cite{sivers}.  Integration over all particles detected
in an acceptance that is azimuthally symmetric around the thrust axis
ensures cancellation of possible contributions from spin- and
$k_T$-dependent fragmentation functions that serve to analyze quark
polarizations transferred to the final state (Collins effect\cite{collins}).  
Particularly for events at large $x_F$, the forward
$\pi^0$ carries a large fraction of the energy of the forward
scattered parton.  A precise definition of jet-like behavior is
required.

We know that jet-like events are present at large $\eta$ from results
with the STAR Forward Pion Detector (FPD).  Fig. \ref{jet_like} shows
the reconstructed invariant 
mass distribution, where
$M_{\gamma\gamma}=E_{trig}\sqrt{1-z_{\gamma\gamma}^2}{\rm
sin}(\phi_{\gamma\gamma}/2)$.  The total energy ($E_{trig}$)
corresponds to the sum of energy from all towers of one of the FPD
modules and is taken as the $\pi^0$ energy in the analysis.  It is
used in conjunction with the opening angle ($\phi_{\gamma\gamma}$) and
energy sharing, $z_{\gamma\gamma}= |(E_{\gamma 1} - E_{\gamma 2}| /
(E_{\gamma 1}+E_{\gamma 2})$, from the two highest energy photons
reconstructed in the event.  Jet-like events occur when more than two
photons are found, resulting in $E_{trig}>E_{\pi}$ and therefore
$M_{\gamma\gamma}>M_{\pi}$.  This is observed in Fig. \ref{jet_like}
and is accounted for by simulation, which is decomposed
into its various contributions.  But, events in the FPD at a given
$x_F$ occur primarily in portions of the calorimeter closest to the
beam because this minimizes $p_T$ (see right panel of
Fig. \ref{STAR_sig_results}).   For such events the FPD does not have
azimuthally symmetric acceptance around the thrust axis for 
additional particles distributed around the reconstructed $\pi^0$.

\begin{figure}[ht]
\centerline{\epsfxsize=2.0in\epsfbox{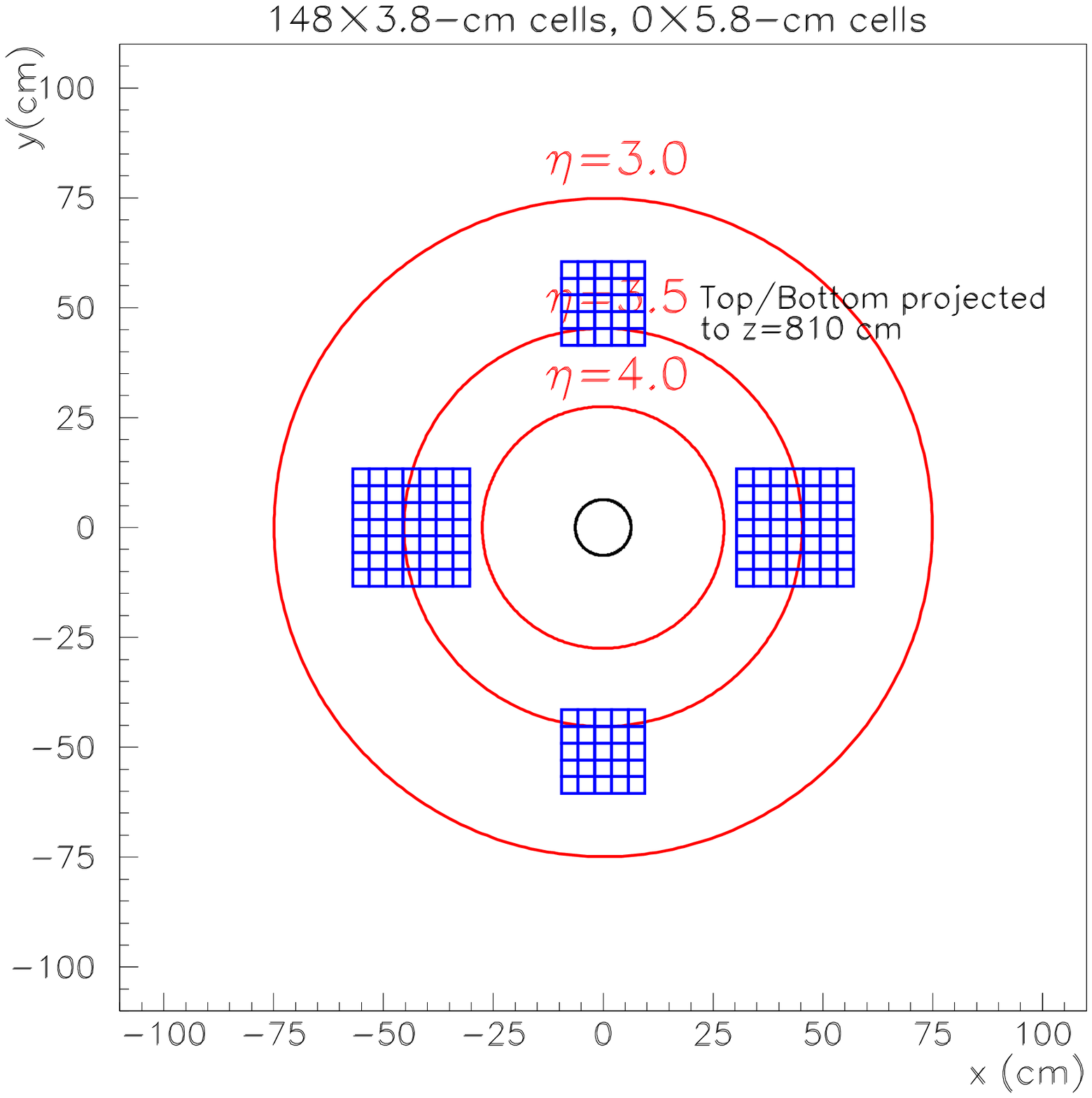}~~\epsfxsize=2.0in\epsfbox{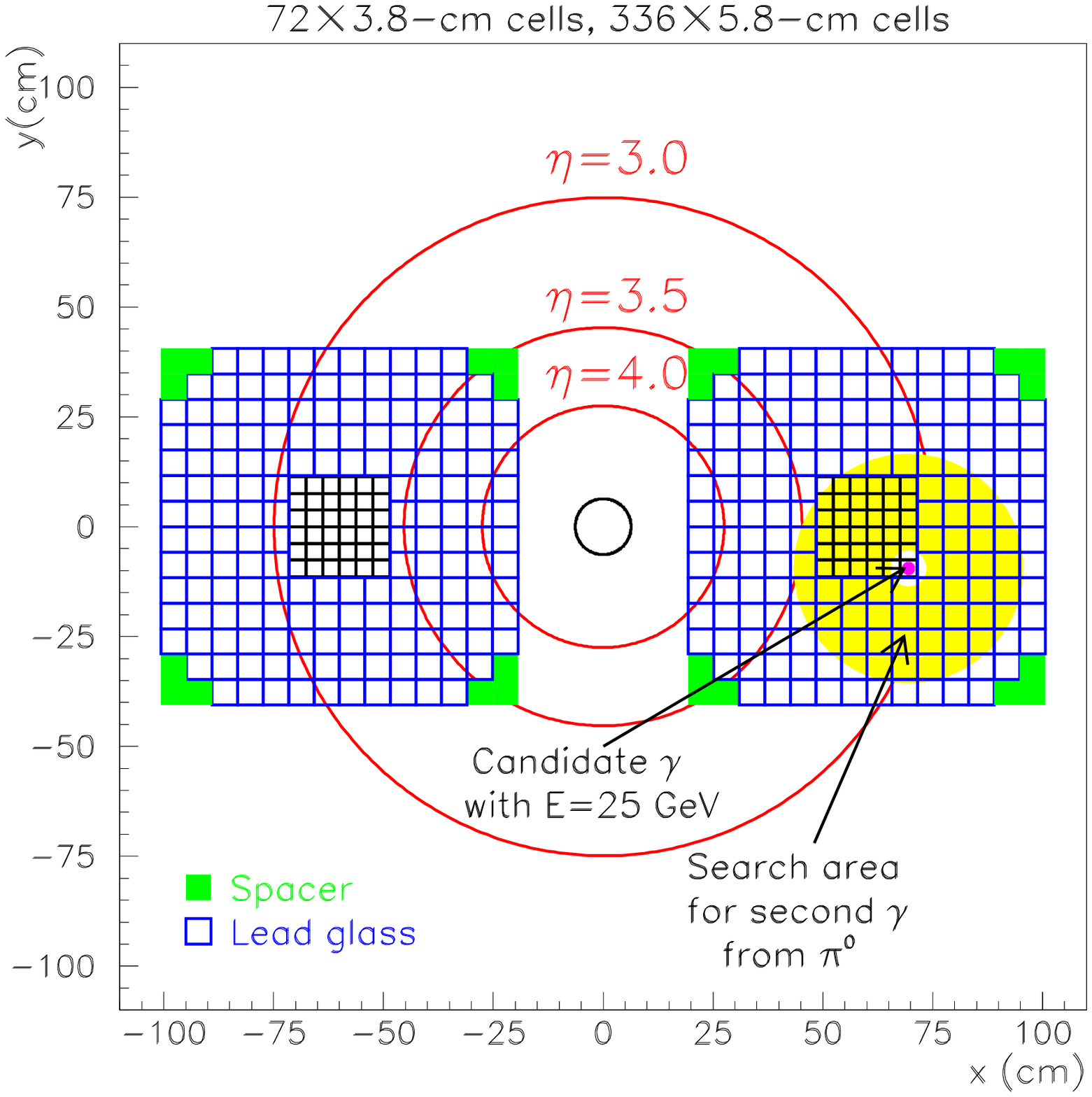}}
\caption{(left) Layout of STAR forward pion detector used in run 5. 
(right) Layout of STAR FPD++ that is planned for use in run 6.\label{STAR_FPD++}}
\end{figure}

The issue of azimuthally symmetric coverage for jet-like events 
is resolved by an upgrade known as the FPD++ (Fig. \ref{STAR_FPD++}) 
that has been built for the upcoming RHIC run as an engineering
test of the STAR Forward Meson Spectrometer (FMS)\cite{FMS}.  It consists of
two left/right symmetric calorimeters that replace the FPD modules
west of the STAR interaction point.  The original FPD modules remain
on the east side of STAR and are planned to improve the precision of
transverse SSA measurements at large $x_F$.  As shown in Fig. \ref{STAR_FPD++}, the inner
portion of each calorimeter module is essentially identical to the
FPD.  The outer portion of the calorimeter consists of larger cells\cite{focus}
that are placed with azimuthal symmetry about the inner portion.  Events can
be selected with the FPD++ in an identical manner as used for the FPD
that result in sizeable transverse SSA for $\pi^0$ production.  The additional detector
coverage can be queried for evidence of additional photons that
accompany a trigger $\pi^0$ thereby signaling jet-like events.  Based on 
the diphoton invariant mass distribution and the photon multiplicity
distribution, at least 16\% of the $\pi^0$ events observed in the FPD
with $E_{\pi}\ge 20 $ GeV are accompanied by additional photons in
jet-like events.  

\begin{figure}[ht]
\centerline{\epsfxsize=2.8in\epsfbox{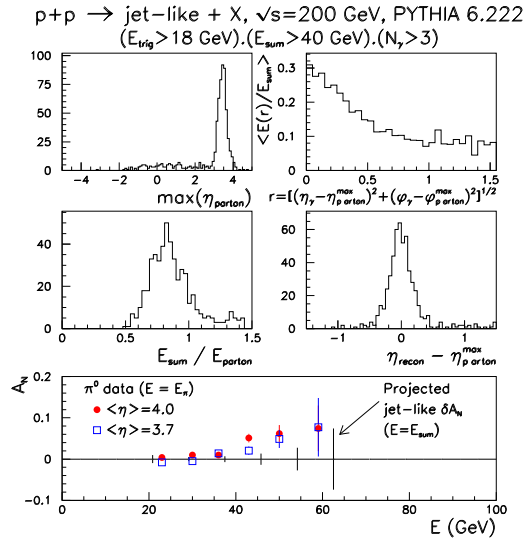}}
\caption{PYTHIA simulations of FPD++ response for p+p collisions at $\sqrt{s}$=200 GeV.  
$E_{trig}$ ($E_{sum}$) is the energy sum in the central section (entire) of the calorimeter
(Fig. \ref{STAR_FPD++}.  (upper left) most forward hard-scattered parton 
pseudorapidity distribution; (upper right) distribution of 
photon energy relative to the thrust axis showing a standard 
jet shape atop an underlying event contribution; (middle left) photon energy 
sum scaled by the most forward hard-scattered parton energy; (middle right)
difference distribution of the $\eta$ reconstructed from the 
vector sum of the detected photon momenta and the parton $\eta$;
(bottom) projected statistical precision for jet-like events for 5 /
pb of polarized proton integrated luminosity. \label{jet_like_sim}}
\end{figure}

Fig. \ref{jet_like_sim} shows PYTHIA 6.222 \cite{pythia} simulations
that provide an operational definition of what we mean by jet-like
events.  PYTHIA is expected to have predictive power in this
kinematics because it has been previously shown to agree with
measured forward pion cross sections\cite{dubna}.  To explore jet-like
events, minimum-bias PYTHIA events are selected when the summed
photon energy in the inner portion of a FPD++ module, defined as
$E_{trig}$, exceeds 18 GeV.  To facilitate possible reconstruction of
forward $\pi^0+\pi^0$ pairs, events are further required to have more
than 3 photons within the full acceptance of an FPD++ module.  These
requirements mean that selected events with energy from incident
photons summed over the entire FPD++ module ($E_{sum}$) may exhibit
jet-like behavior.  The upper left panel of Fig. \ref{jet_like_sim}
shows the pseudorapidity distribution of the most forward angle
hard-scattered parton when $E_{sum}>40$ GeV.  It is peaked at $\eta\approx$3.3,
corresponding to the location of the triggering portion of the FPD++
module.  The small background near midrapidity has contributions from
large Bjorken $x$ quarks that emit initial-state radiation that
subsequently scatters from soft gluons from the other proton.  The
distribution of the photon energy relative to the thrust axis of the
forward scattered parton is shown in the upper right panel of
Fig. \ref{jet_like_sim}.  Jet-like behavior is evident, although
evidence for contributions from the underlying event is also present.
The summed photon energy within the FPD++ acceptance gives a good
representation of the forward scattered parton, albeit shifted in its
energy scale.  Furthermore, the vector sum of the detected photon
momenta faithfully reconstructs ($\eta_{recon}$) the direction of the
scattered parton.  From the middle right panel of
Fig. \ref{jet_like_sim}, the symmetry of the
$\delta\eta=\eta_{recon}-\eta_{parton}^{max}$ distribution indicates
that the underlying event is not skewed from fragments of the beam
jets.  Projections for the uncertainties that could be measured on the
$A_N$ for these jet-like events with 5 pb$^{-1}$ of integrated luminosity
recorded in a data sample with beam polarization of 50\% are shown in
the bottom panel of Fig. \ref{jet_like_sim}.

In addition, the FPD++ is expected to allow for robust detection of
large $x_F$ direct photon production.  Rather than searching for
coincident photons, the outer portion of the calorimeters can serve to
remove neutral meson contributions, to extract events with only a
single energetic photon observed in the calorimeter.  The resulting
yield would be predominantly prompt photon events, including direct
photons and fragmentation photons.  The calorimetric coverage will
allow suppression of most of the latter events.  We expect
$\approx$70,000 direct photon events with $E\ge25$ GeV in a data
sample from the FPD++ that records 5 pb$^{-1}$ of $p+p$ collisions at
$\sqrt{s}$=200 GeV.  These photon energies are larger than the
simulated cutoff in the distribution of energy deposition by 
charged hadrons incident on the calorimeter.

The left/right symmetry of the FPD++ is important for the cancellation
of systematic errors.  With that symmetry, so-called cross ratio
methods can be used for extracting single-spin asymmetries for $\pi^0$
production, jet-like events and prompt photon events.  Another benefit
of the symmetry is for coincident events.  The opening angle between
the two calorimeters is well matched to that required for large-$x_F$
production of objects with invariant mass of order 3 GeV/c$^2$ that
decay to either photons, neutral mesons or electron-positron pairs.

The end result is that the upcoming RHIC run should provide exciting
results for transverse spin physics.  Perhaps of greatest interest is
the prospect for isolation of the Sivers function both at midrapidity
and for forward particle production.  Its isolation may establish
the dynamical origin of transverse SSA for large-$x_F$ $\pi$
production and may conclusively demonstrate orbital motion of the
constituents of the proton.

\section{Acknowledgements}
The forward calorimetry work is done with the STAR collaboration.  
I would like to acknowledge all members of the RHIC spin collaboration
(RSC) for making the RHIC-spin program a reality. The RSC is a
group including members of the RHIC experiments, members of
the BNL Collider-Accelerator Department, experts in the RHIC
polarimeters required to determine the beam polarization, and
theorists.  I would also like to thank Akio Ogawa, Hank Crawford, Jack
Engelage, Carl Gagliardi, Steve Heppelmann, Larisa Nogach, Greg
Rakness, Gerry Bunce and Werner Vogelsang for their comments and help.
This work was supported by DOE Medium Energy Physics.

\end{document}